\documentclass[aps, twocolumn, superscriptaddress,longbibliography, amsfonts,amssymb,amsmath,a4paper,floats,floatfix]{revtex4-2}

\usepackage[utf8]{inputenc}
\usepackage[english]{babel}

\usepackage{graphicx}
\usepackage{dcolumn}
\usepackage{bm}
\usepackage{siunitx}
\usepackage{xspace}
\usepackage{color}

\newcommand{\mos}{MoS\ensuremath{_{\mathrm{2}}}\xspace}
\newcommand{\vtg}{\ensuremath V_{\mathrm{tg}}}
\newcommand{\vbg}{\ensuremath V_{\mathrm{bg}}}   
\begin{document}

\title{Electron transport in dual-gated three-layer \mos}

\author{Michele Masseroni}%
    \affiliation{Solid State Physics Laboratory, ETH Z\"urich, 8093 Z\"urich, Switzerland}
\author{Tim Davatz}%
    \affiliation{Solid State Physics Laboratory, ETH Z\"urich, 8093 Z\"urich, Switzerland}
\author{Riccardo Pisoni}%
    \affiliation{Solid State Physics Laboratory, ETH Z\"urich, 8093 Z\"urich, Switzerland}
\author{Folkert K. de Vries}
    \affiliation{Solid State Physics Laboratory, ETH Z\"urich, 8093 Z\"urich, Switzerland}
\author{Peter Rickhaus}
    \affiliation{Solid State Physics Laboratory, ETH Z\"urich, 8093 Z\"urich, Switzerland}
\author{Takashi Taniguchi}
	\affiliation{International Center for Materials Nanoarchitectonics,  1-1 Namiki, Tsukuba 305-0044, Japan}
\author{Kenji Watanabe}
	\affiliation{Research Center for Functional Materials, 1-1 Namiki, Tsukuba 305-0044, Japan}
\author{Vladimir Fal'ko}
    \affiliation{National Graphene Institute, University of Manchester, Booth St. E. Manchester M13 9PL, United Kingdom}
\author{Thomas\ Ihn}
    \affiliation{Solid State Physics Laboratory, ETH Z\"urich, 8093 Z\"urich, Switzerland}
\author{Klaus\ Ensslin}
    \affiliation{Solid State Physics Laboratory, ETH Z\"urich, 8093 Z\"urich, Switzerland}

\date{\today}

\begin{abstract}
   The low-energy band structure of few-layer \mos is relevant for a large variety of experiments ranging from optics to electronic transport. Its characterization remains challenging due to complex multi band behavior.
   We investigate the conduction band of  dual-gated three-layer \mos by means of magnetotransport experiments. The total carrier density is tuned by voltages applied between \mos and both top and bottom gate electrodes.
   For asymmetrically biased top and bottom gates, electrons accumulate in the layer closest to the positively biased electrode.
   In this way, the three-layer \mos can be tuned to behave electronically like a monolayer. In contrast, applying a positive voltage on both gates leads to the occupation of all three layers.
   Our analysis of the Shubnikov--de Haas oscillations originating from different bands lets us attribute the corresponding carrier densities in the top and bottom layers.
   We find a twofold Landau level degeneracy for each band, suggesting that the minima of the conduction band lie at the $\pm K$ points of the first Brillouin zone. This is in contrast to band structure calculations for zero layer asymmetry, which report minima at the $Q$ points.
   Even though the interlayer tunnel coupling seems to leave the low-energy conduction band unaffected, we observe scattering of electrons between the outermost layers for zero layer asymmetry. The middle layer remains decoupled due to the spin-valley symmetry, which is inverted for neighboring layers.
   When the bands of the outermost layers are energetically in resonance, interlayer scattering takes place, leading to an enhanced resistance and to magneto-interband oscillations.

\end{abstract}

\maketitle

\section{Introduction}

The band structure of semiconducting transition metal dichalcogenides (TMDCs) has been extensively studied from a theoretical \cite{li_electronic_2007, zahid_generic_2013,brumme_first-principles_2015, kormanyos_kp_theory_2015, dias_band_2018} and experimental \cite{finteis_occupied_1997, splendiani_emerging_2010, mak_atomically_2010,han_band-gap_2011, movva_tunable_2018, larentis_large_2018, pisoni_absence_2019, pisoni_interactions_2018} point of view.
The results obtained were sometimes conflicting regarding the location of the valence band maximum and the conduction band minimum in the Brillouin zone.
Optical experiments on molybdenum disulfide (\mos) performed by Splendiani \textit{et al.} \cite{splendiani_emerging_2010} and Mak \textit{et al.} \cite{mak_atomically_2010} demonstrated a band structure transition as a function of the number of \mos layers.
Both reported a transition from an indirect to a direct band gap when the material was thinned down to a single layer.
According to density functional theory (DFT) calculations of the electronic band structure, the conduction band minima of monolayer \mos lie at the $\pm K$ points of the first Brillouin zone \cite{kormanyos_monolayer_2013, kormanyos_kp_theory_2015, brumme_first-principles_2015}.
This result is in agreement with the observation of a twofold Landau level (LL) degeneracy in magnetotransport experiments \cite{pisoni_interactions_2018}.
Due to strong spin-orbit (SO) interaction and a lack of inversion symmetry, the spin degeneracy is lifted \cite{kormanyos_monolayer_2013}, and the observed twofold degeneracy was attributed to the valley degeneracy. This confirmed the theoretically predicted position of the conduction band minima at the $\pm K$ points. 

In few-layer \mos, the interlayer tunnel coupling is predicted to lead to band hybridization, shifting the conduction band at the $Q$ points (midway along the $\Gamma - K$ lines in the first Brillouin zone) downward in energy, whereas the bands at the $K$ valleys are almost unaffected \cite{ellis_indirect_2011}. 
However, DFT calculations do not agree on the size of the energy gaps at the $K$ and $Q$ points. 
While previous calculations showed that the conduction band minima of a bilayer were at the $Q$ points \cite{splendiani_emerging_2010}, more recent calculations found the minima still at the $K$ points \cite{zahid_generic_2013, kormanyos_tunable_2018}.
The observation of a twofold LL degeneracy in bilayer \mos \cite{pisoni_absence_2019} confirmed these recent results.
Furthermore, in magnetotransport experiments performed by Pisoni \textit{et al.} \cite{pisoni_absence_2019}, no evidence of interlayer tunnel coupling was observed, which is the mechanism expected to shift the $Q$ valleys downward in energy.

Angle-resolved photoemission spectroscopy \cite{jin_direct_2013} revealed a shift of the valence band maximum from the $K$ points to the $\Gamma$ point with increasing layer number, which explains the direct-to-indirect band gap transition observed in optical experiments \cite{splendiani_emerging_2010, mak_atomically_2010}.
Theoretical calculations of the band structure of bulk \mos (four and more layers) agree on the position of the conduction band minima at the $Q$ points. 
Therefore, the question arises whether the lowest conduction band minima have changed from the $K$ to the $Q$ valleys already for three-layer (3L) \mos.
There are only few reports \cite{wu_evenodd_2016, pisoni_gate-defined_2017} claiming the experimental observation of $Q$-valley electrons in 3L \mos.
However, band structure calculations of Brumme \textit{et al.} \cite{brumme_first-principles_2015}, which take electron doping via the field effect into account, show that the lowest conduction band minimum in 3L \mos is still located at the $K$ points. To populate the $Q$ valleys, an electron doping $>\SI{2E13}{cm^{-2}}$ is required.

Here, we investigate magnetotransport in a dual-gated 3L \mos device.
We demonstrate that its electronic properties are like a monolayer when we apply different voltage polarities on the two gates.
By applying a positive voltage on both gates, electrons populate multiple bands in multiple layers.
Screening effects between the layers play a major role for the layer population, leading to an almost independent electron density tunability in the outermost layers by electrostatic gating on both sides of the device.
From the analysis of the Shubnikov--de Haas (SdH) oscillations, we determine a twofold band degeneracy and conclude that the lowest conduction band minimum of 3L \mos lies at the $K$ points of the first Brillouin zone.
We observe signatures of interlayer scattering whenever the conduction bands in the two outermost layers are energetically in resonance.
The opening of a scattering channel appears as a peak in the resistance at zero magnetic field, evolving along the zero displacement field line (same density in the outermost layers).
The interlayer scattering mechanism is confirmed by the presence of magneto-interband oscillations (MISOs).
Interlayer scattering has not been observed in bilayer \mos \cite{pisoni_absence_2019} because adjacent layers in the polytype 2H-\mos are rotated by \SI{180}{\degree}. Thus, elastic scattering from one layer to the other would require a spin flip or a momentum transfer on the scale of the size of the first Brillouin zone, both of which are unlikely to happen.

\section{Results and discussion}

\subsection{Sample and measurement technique}

\begin{figure}[tb]
\begin{center}
\includegraphics[scale=1]{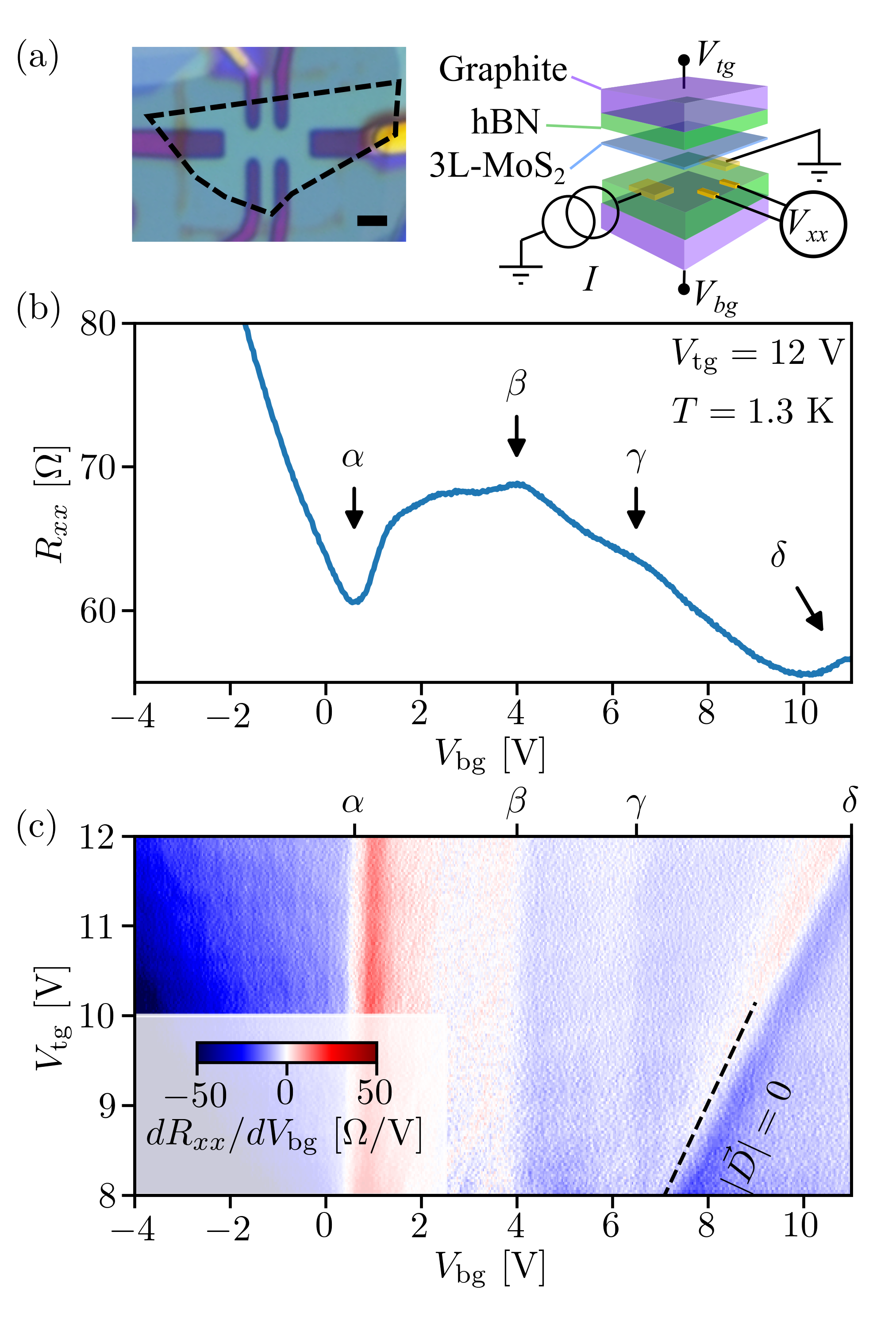}
\caption{(a) Optical image of the sample (left panel). The three-layer (3L) \mos flake is outlined by the dashed lines. 
The scale bar has a length of $\SI{3}{\mu m}$.
Schematic view of the device (right panel): 3L \mos encapsulated in hexagonal boron nitride (hBN) with metallic bottom contacts, bottom and top graphite gates. 
(b) Resistance $R_{xx}$ as a function of $\vbg$ at $\vtg=\SI{12}{V}$ and $T=\SI{1.3}{K}$.
The arrows (labeled $\alpha$, $\beta$, $\gamma$, $\delta$) point to features  where $R_{xx}(V_\mathrm{bg})$ does not simply follow a smooth dependence dictated by sheet density increase with bottom gate voltage.
(c) Numerical derivative of $R_{xx}$ taken along the voltage axis $\vbg$ plotted as a function of $\vbg$ and $\vtg$.
The tilted dashed line shows a line of constant displacement field.}
\label{Fig:Fig1}
\end{center}
\end{figure}

To investigate electronic transport in 3L \mos, we employed a dual-gated field-effect transistor with metallic ohmic contacts on the substrate side.
An optical image and a schematic side view of our device are shown in Fig.~\ref{Fig:Fig1}(a).
The heterostructure was assembled on a silicon/silicon-oxide substrate using a polymer-based dry transfer technique.
We employed graphite gates and used hexagonal boron nitride (hBN) as the gate dielectric.
The thicknesses of the bottom and top hBN layers are $\SI{19.4}{nm}$ and $\SI{22.3}{nm}$, respectively. 
The \mos flakes were obtained by mechanically cleaving a bulk crystal of natural sources (SPI supplies). The exfoliation and stacking was carried out inside a glove box with argon atmosphere ($\mathrm{O_2, H_2O}<\SI{0.1}{ppm}$).
The number of \mos layers was identified by optical contrast and confirmed by atomic force microscopy only after encapsulation. The fabrication process can be divided into four main steps: assembly of the bottom graphite  and hBN layers, deposition of the metal contacts (Ti/Au: \SI{5}{nm}/\SI{15}{nm}), cleaning the contact region with the tip of a scanning force microscope in contact mode, and assembly of the top graphite-hBN-\mos heterostructure.
Our fabrication technique allows us to reliably obtain contacts with ohmic behavior at cryogenic temperatures, if we apply a sufficiently large top-gate voltage (see Ref.~\cite{pisoni_interactions_2018}). 
The contact resistance is lower than $\SI{5}{k\Omega}$ for $\vtg\geq \SI{8}{V}$ and reaches $\sim\SI{1}{k\Omega}$ at $\vtg=\SI{12}{V}$. 
In our experiments we applied a constant alternating current bias $I$ ($\SI{100}{nArms}$) and measured the four-terminal longitudinal voltage $V_{xx}$ with standard low-frequency ($\sim\SI{30}{Hz}$) lock-in technique.
The measurement scheme is depicted in Fig.~\ref{Fig:Fig1}(a).

\subsection{Field effect}

\begin{figure*}[tb]
\begin{center}
\includegraphics[scale=1]{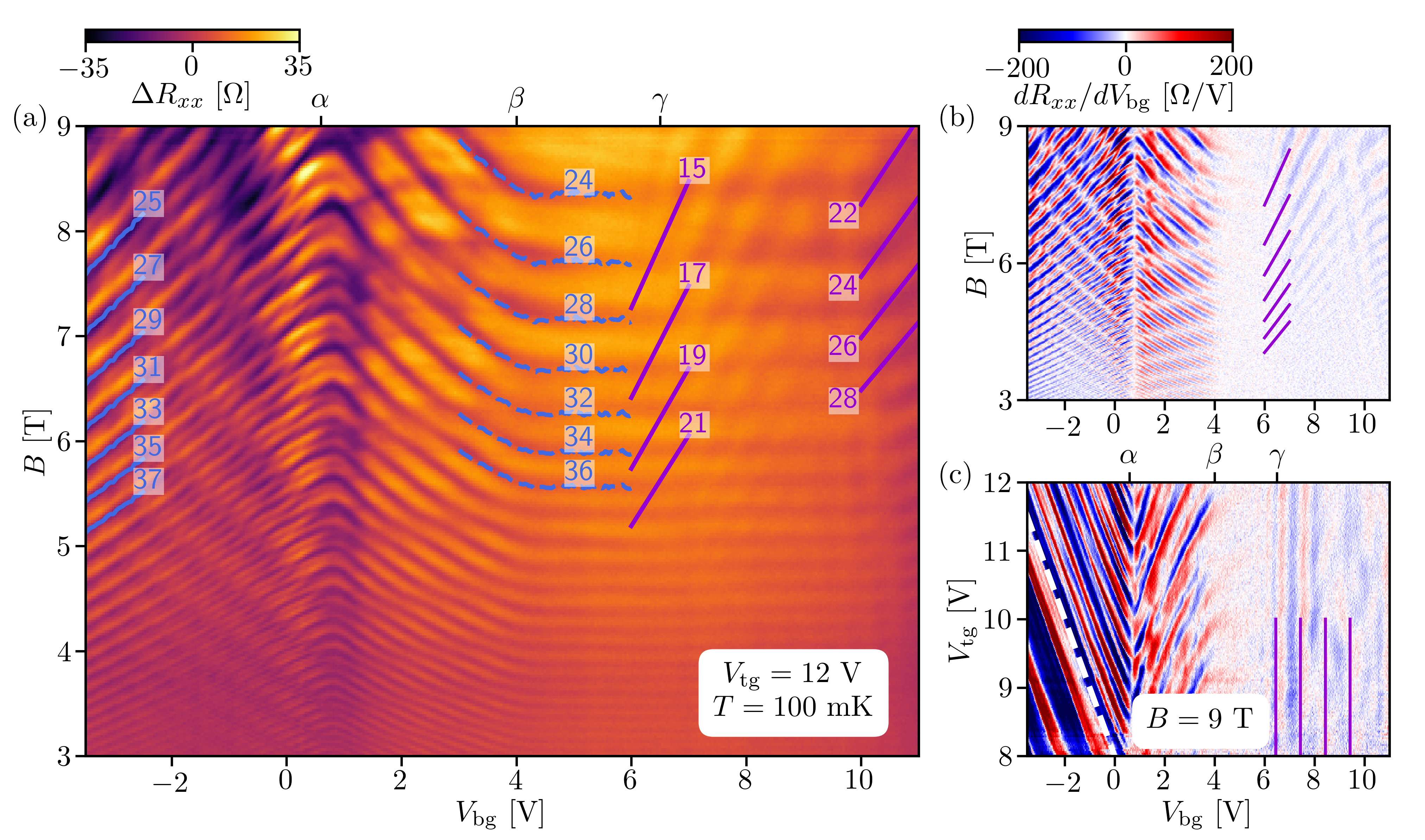}
\caption{(a) Magnetoresistance $\Delta R_{xx}(B,\vbg) = R_{xx}(B,\vbg) - R_{xx}(0,\vbg)$ as a function of $B$ and $\vbg$ for $\vtg=\SI{12}{V}$ and $T=\SI{100}{mK}$. 
The blue and violet lines indicate the positions of the minima of $\Delta R_{xx}$. 
The corresponding numbers refer to the filling factors $\nu_1=hn_1/eB$ (blue) and $\nu_3=hn_3/eB$ (violet) of the top and bottom layers, respectively.
(b) Numerical derivative $dR_{xx}/dV_\mathrm{bg}$ of (a).
(c) $dR_{xx}/dV_\mathrm{bg}$ as a function of $\vtg$ and $\vbg$ at $B=\SI{9}{T}$.
The white dashed line shows a line of constant electron sheet density. }
\label{Fig:Fig2}
\end{center}
\end{figure*}

Figure~\ref{Fig:Fig1}(b) presents the resistance $R_{xx}=V_{xx}/I$ as a function of bottom-gate voltage $\vbg$ at the constant top-gate voltage $\vtg=\SI{12}{V}$, measured at the temperature $T=\SI{1.3}{K}$.
The resistance shows a nonmonotonic behavior as a function of $\vbg$, where we highlight four characteristic gate voltages.
The most prominent feature ($\alpha$) is the onset of an increase in resistance upon increasing $\vbg$ (and thereby total density) at $\vbg^{(\alpha)} \approx  \SI{0.6}{V}$.  
Furthermore, we observe two bumps ($\beta$, $\gamma$) at $\vbg^{(\beta)} \approx  \SI{4}{V}$ and $\vbg^{(\gamma)} \approx  \SI{6.5}{V}$, respectively. 
An additional resistance increase ($\delta$) occurs close to $\vbg^{(\delta)} \approx  \SI{11}{V}$.

To enhance the visibility of these small features, we plot $dR_{xx}/dV_\mathrm{bg}$ in a top-gate--back-gate map in Fig.~\ref{Fig:Fig1}(c).
We find that the features labeled $\alpha$, $\beta$, and $\gamma$ are independent of $\vtg$. Feature $\delta$ is not; it follows the line of zero displacement field  
\begin{equation}
   D= \frac{C_\mathrm{tg} V_\mathrm{tg} - C_\mathrm{bg} V_\mathrm{bg}}{2\epsilon_0}, 
\end{equation}
with $C_\mathrm{tg}$, $C_\mathrm{bg}$ being the geometric gate capacitances per area between the respective gate and the 3L \mos, and $\epsilon_0$ being the vacuum permittivity.

Concerning feature $\alpha$ in Fig.~\ref{Fig:Fig1}(b), we attribute the large increase in resistance  to increasing electron scattering when another layer (or band) becomes populated.
The enhanced electron scattering reduces the mobility and thereby leads to an increase in resistance, even though a new conducting channel becomes available for transport.
Similar drops in mobility have been observed in semiconductor quantum wells when occupying a second subband \cite{stormer_observation_1982, ensslin_single-particle_1993, tschirky_scattering_2017}.
The independence on top gate voltage in Fig.~\ref{Fig:Fig1}(c) indicates that the action of the top gate on feature $\alpha$ is screened by a charged layer between the top gate and the newly populated layer or band. This feature can therefore only be associated with the onset of the occupation of a band in the middle or lower layer of the structure for $\vbg>\vbg^{(\alpha)}$. Inversely, for $\vbg<\vbg^{(\alpha)}$, conduction must take place in either the top or the middle layer.

The smaller bumps in the resistance ($\beta$, $\gamma$) may have the same origin as $\alpha$. 
The three features in the resistance ($\alpha$, $\beta$, $\gamma$) share the property of being independent of $\vtg$ and can therefore only be attributed to the occupation of bands in the middle or the lowest layer. 

\begin{figure}[tb]
\begin{center}
\includegraphics[scale=1]{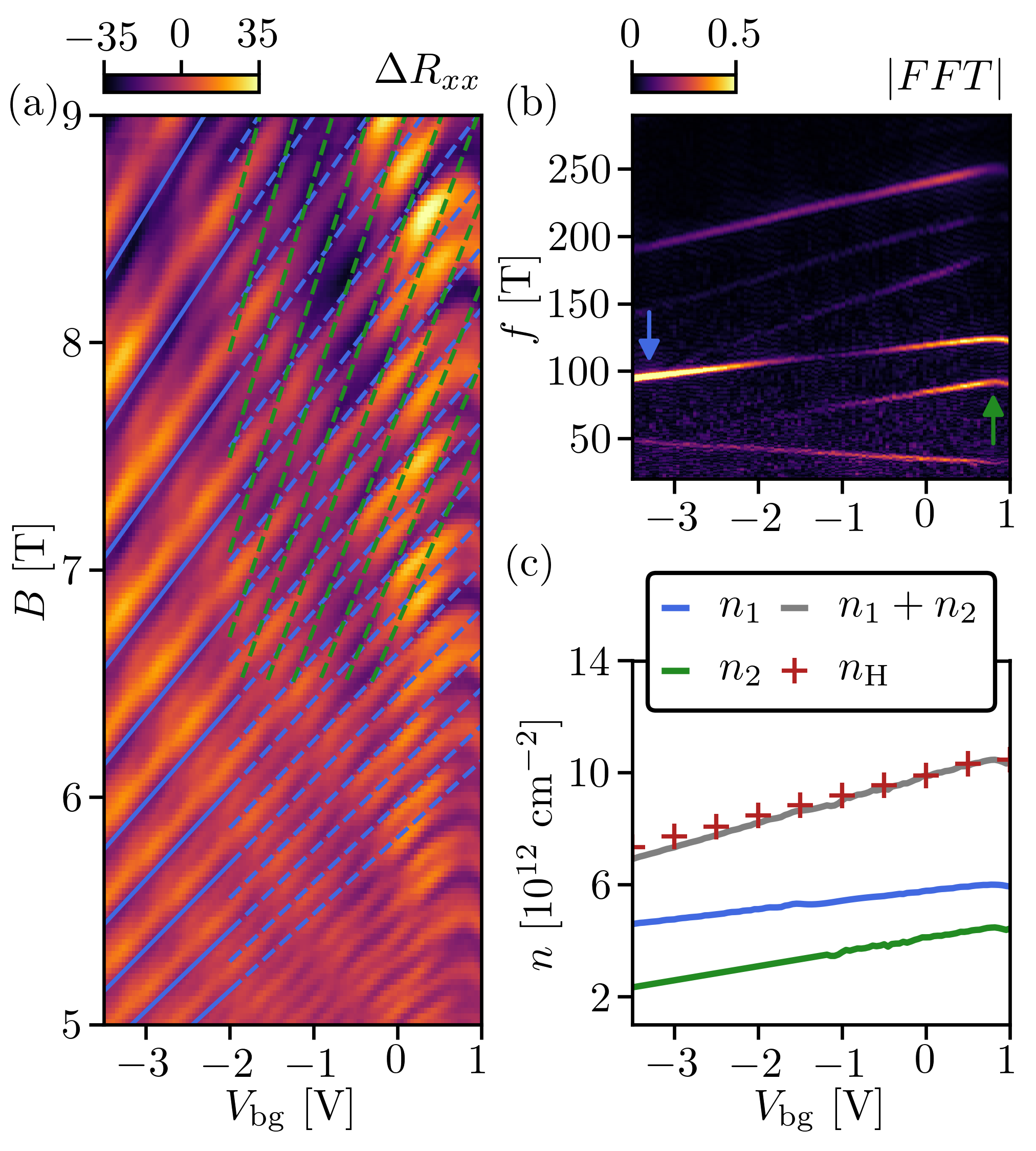}
\caption{(a) Zoom-in of Fig.~\ref{Fig:Fig2}(a). 
The blue (green) lines represent filling factors $\nu_1=eB/hn_1$ ($\nu_2=eB/hn_2$) of band 1 (2).
(b) Amplitude spectrum of the fast Fourier transform (FFT) of $\Delta R_{xx}(1/B)$.
The blue (green) arrow marks the frequency $f_1$ ($f_2$).
Density as a function of $\vbg$ assuming twofold degenerate bands.
The solid lines are obtained from the frequencies of the Fourier spectrum by means of Eq.~\eqref{eq:fftfreq}.
The red crosses are obtained from the slope of the Hall resistance.}
\label{Fig:Fig3}
\end{center}
\end{figure}

\subsection{Magnetoresistance oscillations}

To confirm the hypothesis of enhanced scattering by the occupation of additional bands, we measure the occupation of bands by magnetotransport measurements. We applied a magnetic field $B$ perpendicular to the sample plane and measured $R_{xx}$ as a function of bottom-gate voltage.
The minima of SdH oscillations in $R_{xx}(B)$ occur when $m=hn_i/g_ieB=\nu_i/g_i$ is an integer (with $n_i$ being the electron density, $g_i$ the LL degeneracy, and $\nu_i$ the LL filling factor of band $i$). 
Therefore, a sequence of integer numbers $m$ leads to a set of lines
\begin{equation}
B_m(\vbg) = \frac{hn_i(\vbg)}{eg_im}
\label{eq:lfan}
\end{equation}
in the $(\vbg, B)$ plane corresponding to filling factors $\nu_i=mg_i$, which we hereafter refer to as the Landau fan of band $i$.
In the presence of multiple occupied bands, we expect to observe multiple Landau fans. Note that nonlinearities in the density $n_i(\vbg)$ of band $i$ may lead to nonlinearities in $B_m(\vbg)$, even though the total sheet electron density $n_\mathrm{s}$ increases linearly with gate voltage.

Figure~\ref{Fig:Fig2}(a) shows the magnetoresistance $\Delta R_{xx}$ in the plane $(\vbg, B)$ measured at $T=\SI{100}{mK}$.
The measurement was performed at a constant top-gate voltage ($V_\mathrm{tg}=\SI{12}{V}$) ensuring that the contacts were ohmic.
The bottom gate voltage $\vbg$ was expected and confirmed by the classical Hall resistance to be proportional to the total sheet electron density $n_\mathrm{s}\approx n_\mathrm{H}$ [see Fig.~\ref{Fig:Fig3}(c) and the discussion of $n_\mathrm{H}$ below].
The gate-voltage dependence of the magnetoresistance reveals the presence of two Landau fans (blue and violet lines), which create an intricate, in parts nonlinear pattern.
We indicate on the top of Fig.~\ref{Fig:Fig2}(a) the letters $\alpha$, $\beta$ and $\gamma$, marking the voltages at which the corresponding features appear in Fig.~\ref{Fig:Fig1}(b).
Close to each of these voltages, we observe distinct changes in the magnetoresistance oscillations.
This confirms the intimate relation between the resistance features $\alpha$, $\beta$, and $\gamma$ in Fig.~\ref{Fig:Fig1}(b) and \ref{Fig:Fig1}(c) and the population $n_i(\vbg)$ of bands that can be extracted from the magnetoresistance oscillations in Fig.~\ref{Fig:Fig2}(a) with the help of Eq.~\eqref{eq:lfan}.

Starting from the most negative values of $\vbg$, we observe a Landau fan (blue solid lines) with a positive slope, as expected for electron transport in the conduction band at increasing sheet electron density.
Increasing $\vbg$, the slope of this Landau fan changes sign as we pass $\alpha$. 
The slopes $dB_m/d\vbg$ of the Landau fan in Eq.~\eqref{eq:lfan} are proportional to the derivative $dn_i/d\vbg$ of the band density $n_i$ from which the oscillations originate. 
Therefore, the negative slope indicates a reduction of the density in this band upon increasing $\vbg$ beyond $\vbg^{(\alpha)}$, which results from a charge redistribution between two bands.
Further increasing $\vbg$, we observe beyond $\vbg^{(\beta)}$ a second change of the Landau fan slope, which this time becomes almost zero, indicating a constant density $n_i$ of the band relating to the blue Landau fan.
The independence of the density on the gate voltage is a signature of screening effects. 

On the right of $\gamma$, a new Landau fan arises (violet lines) and overlaps with the first one. It becomes more evident in Fig.~\ref{Fig:Fig2}(b), where we plot the derivative $dR_{xx}/d\vbg$.
The presence of a second Landau fan confirms the population of at least an additional band.

To obtain insight into the sensitivity of the oscillations on both top and bottom gates, we show in Fig.~\ref{Fig:Fig2}(c) the SdH oscillations at the constant magnetic field $B=\SI{9}{T}$ as a function of the top- and bottom-gate voltages. 
For $\vbg<\vbg^{(\alpha)}$, we observe SdH oscillations that depend linearly on both gate voltages.
The ratio between the gate capacitances obtained from a line of constant densities (e.g., white dashed line) is $C_\mathrm{bg}/C_\mathrm{tg}=-\Delta \vtg /\Delta \vbg|_{n=\mathrm{const}}= 1.03(5)$.
This is not in good agreement with the ratio of the hBN layer thicknesses ($d_\mathrm{t}/d_\mathrm{b}=1.13$), suggesting that the conducting plane in which these oscillations occur lies closer to the top gate (i.e., in the top layer).
In this case, the lower two \mos layers act as additional dielectric layers, reducing $C_\mathrm{bg}$ compared with the capacitance of the hBN insulator alone.
This result shows that for strongly positive $\vtg$ and negative $\vbg<\vbg^{(\alpha)}$ electrons accumulate in the top layer only, while the middle and bottom layers remain at zero carrier density.

The occupation of an additional band with electrons for $\vbg>\vbg^{(\alpha)}$ and the concurrent decreasing electron density in the top layer [see Fig.~\ref{Fig:Fig2}(a)] lead to the sign reversal of the slope of the oscillations in Fig.~\ref{Fig:Fig2}(c).
The violet Landau fan seen in Figs.~\ref{Fig:Fig2}(a) and \ref{Fig:Fig2}(a)(b) appears in Fig.~\ref{Fig:Fig2}(c) as vertical lines independent of the top-gate voltage.
This independence must be the consequence of strong population of the top layer with electrons, which screens the influence of the top gate on the lower layers.
Therefore, this Landau fan has to originate from the middle or bottom layer. 
Also, all the transitions in the magneto-oscillation patterns in Fig.~\ref{Fig:Fig2}(c) at $\alpha,\beta$, and $\gamma$ appear to be independent of top-gate voltage, like the corresponding features in Fig.~\ref{Fig:Fig1}(c). 
To further disentangle the sequence of layer and band populations with increasing $\vbg$, in the following, we will perform a detailed Fourier-transform analysis of the magnetoresistance oscillations in the different regimes.

\subsection{Single layer regime}

We start the more detailed discussion by further analyzing the regime, where only the top layer is occupied.
Figure~\ref{Fig:Fig3}(a) magnifies the range $\vbg<\vbg^{(\alpha)}$ of Fig.~\ref{Fig:Fig1}(a).
Here, the observed SdH oscillations exhibit an intricate beating pattern at $V_\mathrm{bg} > \SI{-2}{V} $ and magnetic fields $ B > \SI{6}{T} $ [dashed blue and green lines in Fig.~\ref{Fig:Fig3}(a)].
To unravel the related electron densities, we calculate the fast Fourier transform (FFT) of $\Delta R_{xx}(1/B)$. 
The square root of the resulting power spectral density is plotted against the frequency $f=[\Delta(1/B)]^{-1}$ (where $\Delta(1/B)$ is the period of the SdH oscillations plotted against $1/B$), and $\vbg$ in Fig. \ref{Fig:Fig3}(b).
From Eq.~\eqref{eq:lfan}, we find the well-known relation
\begin{equation} \label{eq:fftfreq}
f_i(\vbg)=\left[\Delta\left(\frac{1}{B}\right)\right]^{-1} = \frac{h}{e}\frac{n_i(\vbg)}{g_i},
\end{equation}
relating the fundamental frequency $f_i$ of band $i$ to its density $n_i$.
The spectrum reveals the presence of two fundamental frequencies $f_1$ and $f_2$ (blue and green arrows), suggesting that two distinct bands contribute to electronic transport.
In the voltage range $V_\mathrm{bg}<\SI{-2}{V}$, the SdH oscillations are governed by the first band of frequency $f_1$ [blue lines in Fig.~\ref{Fig:Fig3}(a)], where the corresponding electron density $n_1$ is significantly larger than in the second band. 
In the voltage range $\vbg>\SI{-2}{V}$, both bands contribute to the oscillations, leading to the beating observed in Fig.~\ref{Fig:Fig3}(a).
At large magnetic fields, a progressive lifting of the LL degeneracy becomes evident and leads to the presence of peaks at $2f_i$ ($i=1,2$) in the spectrum, as the degeneracy factors $g_i$ in Eq.~\eqref{eq:fftfreq} change to half of their previous values. The population of two bands in the top layer makes it plausible that this layer screens the electric field from the top gate completely, such that the features $(\alpha)$, $(\beta)$, and $(\gamma)$ related to the lower layers are completely independent of top-gate voltage.

\begin{figure}[tb]
\begin{center}
\includegraphics[scale=1]{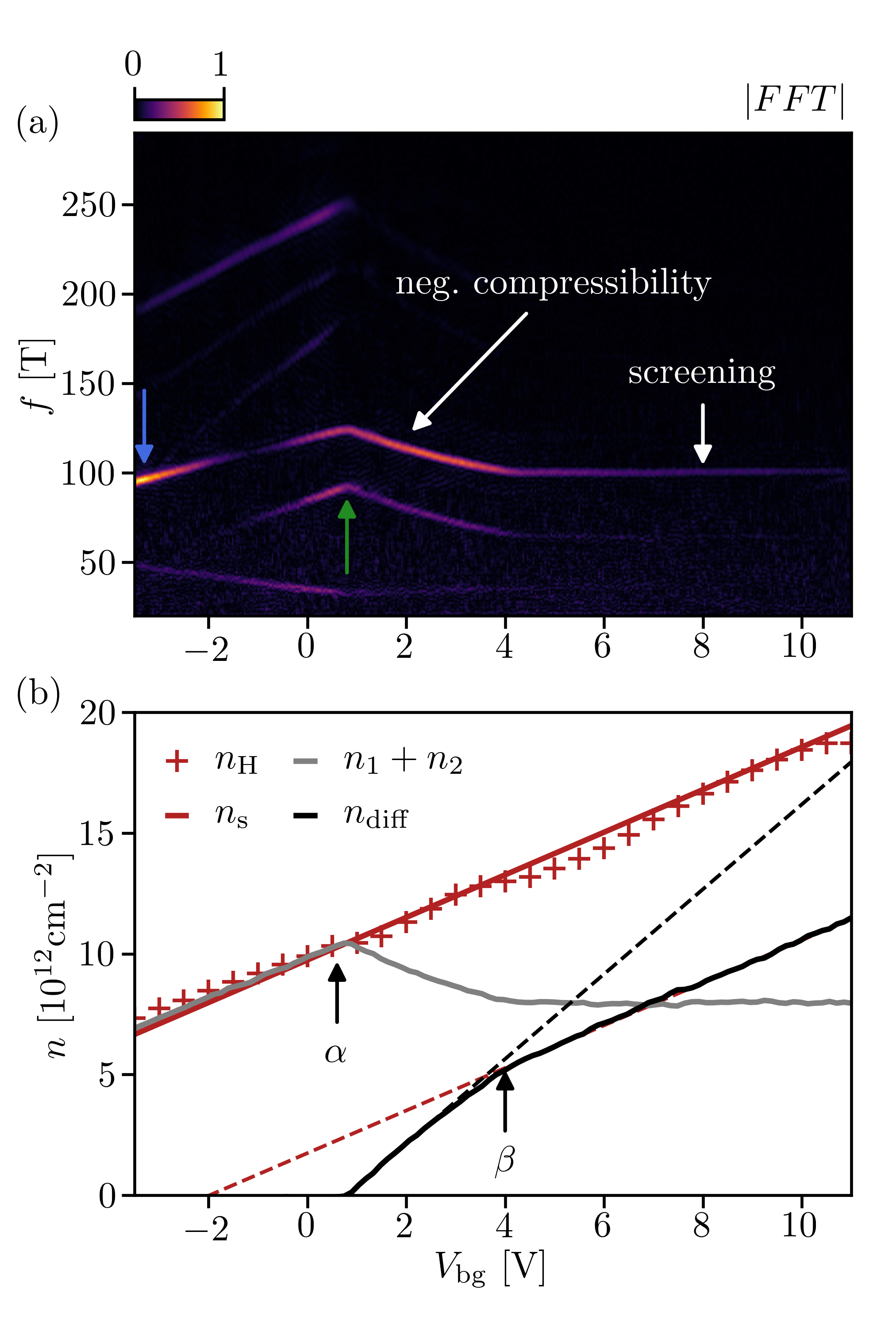}
\caption{(a) Amplitude spectrum of the fast Fourier transform (FFT) of $\Delta R_{xx}(1/B)$ shown in Fig.~\ref{Fig:Fig2}(a). 
(b) Density as a function of $\vbg$. The density $n_1+n_2$ (black dotted line) is obtained from the Shubnikov--de Haas (SdH) oscillation frequencies, while $n_\mathrm{s}$ is calculated with a plate capacitor model.
The density obtained from the slope of the Hall resistance is shown by red crosses. 
The black dashed line corresponds to a linear fit of $n_\mathrm{diff}$ for $\vbg<\SI{4}{V}$.
The red dashed line corresponds to ($n_\mathrm{s}-\SI{8E12}{cm^2}$), where $\SI{8E12}{cm^2}$ corresponds to the saturated density in the top layer.}
\label{Fig:Fig4}
\end{center}
\end{figure}

A standard method for determining the degeneracies $g_i$ of the LLs in particular bands consists of comparing the frequency of the SdH oscillations defined in Eq.~\eqref{eq:fftfreq} with the sheet density $n_\mathrm{H}$ obtained from the slope of the classical Hall resistance ($1/en_\mathrm{H}$).
Our sample has no proper Hall bar geometry and has metallic contacts extending into the conducting channel, which can lead to an overestimation of the sheet density.
Nevertheless, we can consider $n_\mathrm{H}$ as an upper bound for the total sheet density $n_\mathrm{s}$.
The density obtained from the SdH oscillations and the density obtained from the Hall resistance are compared in Fig.~\ref{Fig:Fig3}(c), where we find $n_1+n_2\approx n_\mathrm{H}\approx n_\mathrm{s}$, assuming the degeneracies $g_i=2$ for both bands ($i=1,2$).
This degeneracy $g_{i}=2$ is consistent with the observed halved LL degeneracy at large $B$, which is evident in Fig.~\ref{Fig:Fig3}(a).
We can exclude even larger degeneracy factors because, for $g>2$, the density $n_1+n_2$ would by far exceed $n_\mathrm{H}$.

In the light of the Fourier analysis in Figs.~\ref{Fig:Fig3}(b), \ref{Fig:Fig3}(c) and \ref{Fig:Fig4}(a), now we consider how the densities $n_1$ and $n_2$ evolve as a function of $\vbg$.
Both densities first increase with increasing $\vbg$ up to $\vbg^{(\alpha)}$, followed by a decrease upon increasing $\vbg$, until they finally saturate at $\vbg>\vbg^{(\beta)}$, in complete agreement with our previous discussion of Fig.~\ref{Fig:Fig2}(a). 
Since both densities behave qualitatively similar as a function of $\vbg$, we conclude that both bands belong to the top layer.

This kind of layer polarization was already observed in dual-gated bilayer \mos \cite{pisoni_absence_2019} and in ionic-liquid gated few layer \mos \cite{chen_inducing_2017}, as well as in other bilayer TMDCs \cite{fallahazad_shubnikovhaas_2016, larentis_large_2018}.
Following the results of band structure calculations \cite{li_electronic_2007, zahid_generic_2013,brumme_first-principles_2015, kormanyos_kp_theory_2015, dias_band_2018} and previous experiments \cite{pisoni_interactions_2018, pisoni_absence_2019} we interpret the two bands as the upper and lower SO split bands and define $n_\mathrm{TL} = n_1 + n_2$ as the electron density in the top layer.

Having identified $g_i=2$, the gate capacitance obtained from the slope of $n_\mathrm{TL}$ as a function of  $\vbg$ is $C_\mathrm{bg}=\SI{134}{nF/cm^2}$.
This value is consistent with the value $\SI{139 \pm 5}{nF/cm^2}$ obtained from the parallel plate capacitor model, where we assumed two \mos layers as an additional dielectric material ($\epsilon_\mathrm{hBN}=3.2$ \cite{rickhaus_gap_2019} and $\epsilon_\mathrm{MoS_2}=7$ \cite{laturia_dielectric_2018}).
A reasonable agreement with the experiment is also obtained without considering the empty \mos layers ($\sim \SI{4}{\percent}$ larger capacitance). 
The twofold LL degeneracy suggests that the minimum of the conduction band lies at the $\pm K$ points of the first Brillouin zone, as previously observed in monolayer \cite{pisoni_interactions_2018} and bilayer \mos \cite{pisoni_absence_2019}.
DFT band structure calculations of TMDC-based field effect transistors performed by Brumme \textit{et al.} \cite{brumme_first-principles_2015} are in agreement with our results.
However, our results are in contrast to previous reports on the observation of $Q$-valley electrons in 3L \mos \cite{wu_evenodd_2016, pisoni_gate-defined_2017}.

\begin{figure}[tb]
\begin{center}
\includegraphics[scale=1]{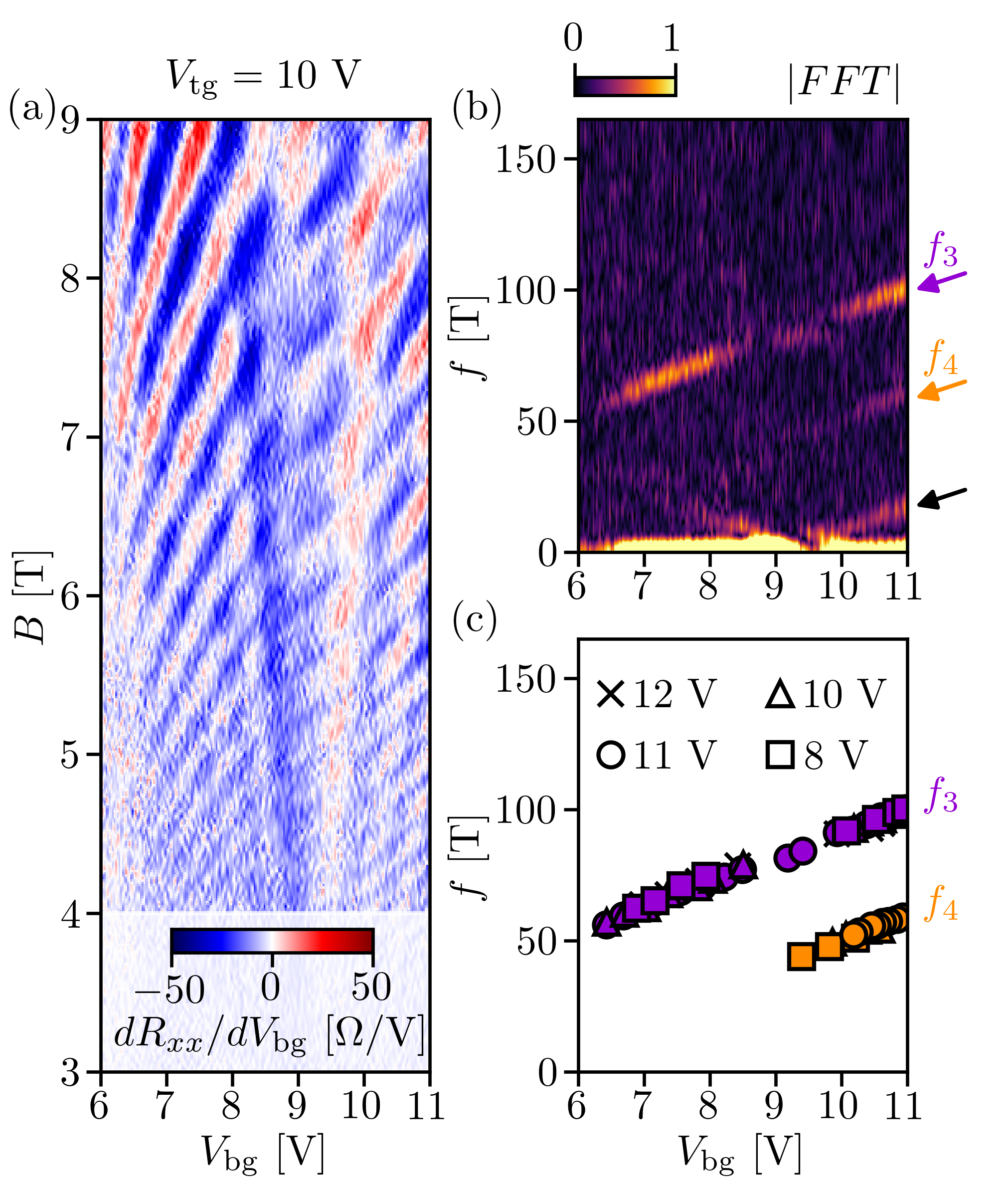}
\caption{(a) $dR_{xx}/dV_\mathrm{bg}$ as a function of $B$ and $\vbg$ at $\vtg=\SI{10}{V}$.
(b) Fast Fourier transform (FFT) of (a). The orange and violet arrows indicates the frequency components of the Shubnikov--de Haas (SdH) oscillations $f_3$ and $f_4$ originating in the bottom layer.
The black arrow indicates a frequency component proportional to the difference of the density in the outermost layers.
(c) Frequencies $f_3$ and $f_4$ obtained for $\vtg$ between $\SI{8}{V}$ and $\SI{12}{V}$.}
\label{Fig:Fig5}
\end{center}
\end{figure}

\subsection{Multi-layer regime}

We now consider the regime where we start populating the lower layers. 
At $V_\mathrm{bg}>\vbg^{(\alpha)}$, the slope of the Landau fan seen in Fig.~\ref{Fig:Fig2}(a) changes sign from positive to negative, indicating a reduction of the density in the top layer at increasing total density as discussed above.
This is also evident from Fig.~\ref{Fig:Fig4}(a), where we show the FFT of the magnetoresistance of Fig.~\ref{Fig:Fig2}(a) over the full range of $\vbg$. 
Both $n_1$ and $n_2$ drop with increasing $\vbg$, corresponding to a dramatic reduction ($\sim \SI{25}{\percent}$) of charge density in the top layer ($n_\mathrm{TL}$), even though the total sheet electron density $n_\mathrm{s}$ keeps increasing with $\vbg$ as shown by the red solid line in Fig.~\ref{Fig:Fig4}(b). 
The total sheet density is calculated with the parallel plate capacitor model, which yields
\begin{equation}\label{eq:Density}
    n_s(\vbg, \vtg) = \frac{C_\mathrm{bg} \vbg + C_\mathrm{tg} \vtg}{e}.
\end{equation}
The result of this calculation is supported by the density $n_\mathrm{H}$ obtained from the Hall effect (red crosses), which steadily increases, showing only slight dips close to the features $\alpha$ and $\beta$. Figure~\ref{Fig:Fig4}(b) also shows $n_\mathrm{TL}$ as the gray solid line exhibiting again the density decrease in the top layer between $\vbg^{(\alpha)}$ and $\vbg^{(\beta)}$.

\begin{figure}[tb]
\begin{center}
\includegraphics[scale=1]{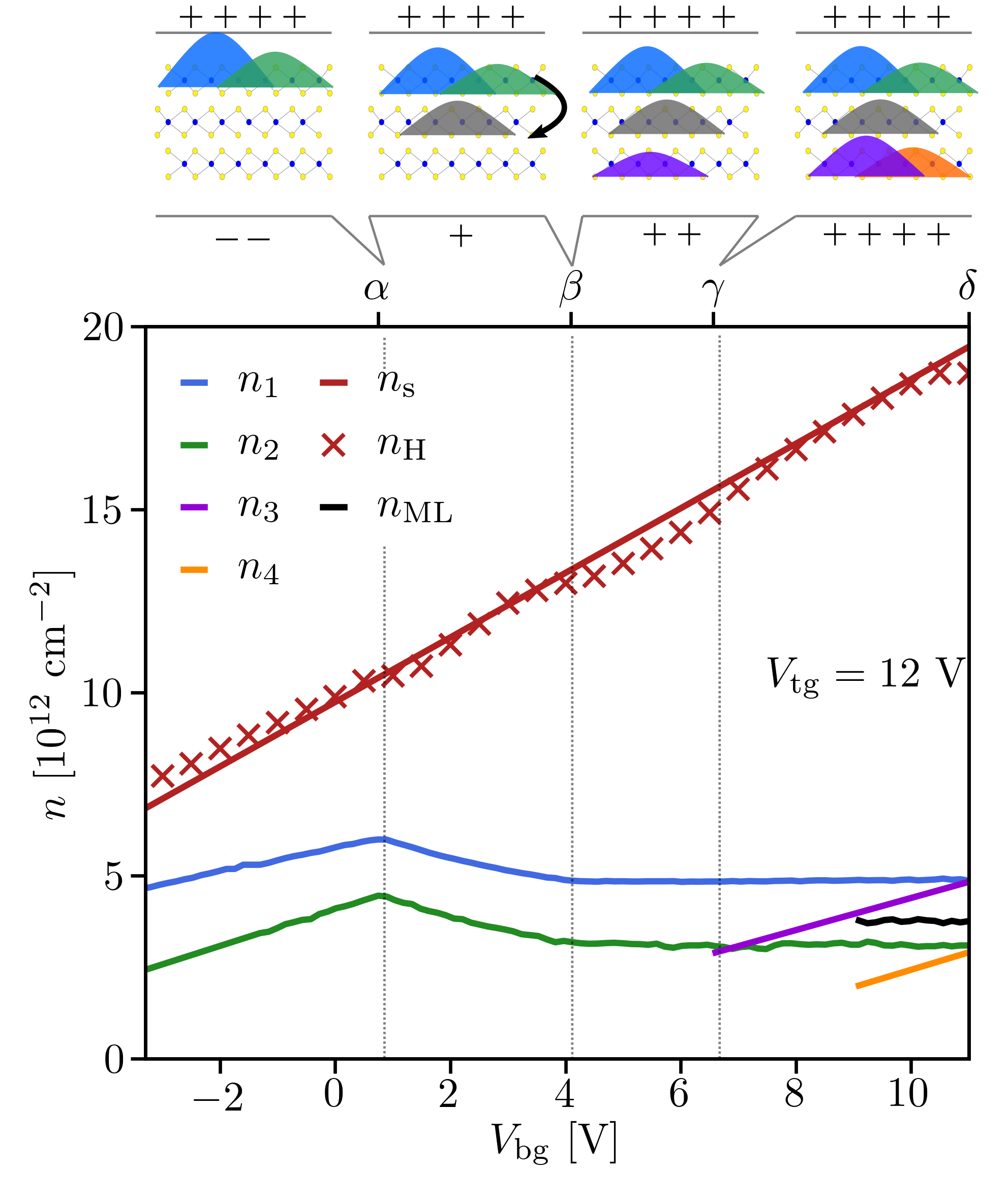}
\caption{Density as a function of $\vbg$ at $\vtg=\SI{12}{V}$. The densities $n_i$ ($i=1,2,3,4$) are obtained from the Fourier analysis of the Shubnikov--de Haas oscillations. $n_\mathrm{s}$ is the total electron sheet density obtained from a parallel plate capacitor model, and $n_\mathrm{H}$ is the density obtained from the slope of the Hall resistance. 
The electron density in the middle layer $n_\mathrm{ML}$ is inferred by charge conservation. 
The top panel shows a schematic of the electron density distribution among the layers. 
The colors used refer to the densities shown below.}
\label{Fig:Fig6}
\end{center}
\end{figure}

A similar density reduction has been observed in bilayer \mos \cite{pisoni_absence_2019} and bilayer WSe$_2$ \cite{fallahazad_shubnikovhaas_2016}.
In the case of bilayer devices, the population of the bottom layer was also accompanied by a reduction of the charge density in the top layer with increasing gate voltage, which was interpreted as a sign of negative compressibility ($K=n^{-1} \frac{dn}{d\mu}$, $\mu$ being the electrochemical potential).
This effect was studied in GaAs double quantum wells \cite{eisenstein_negative_1992} and later also in hybrid graphene-\mos devices \cite{larentis_band_2014} and bulk WSe$_2$ \cite{riley_negative_2015} and it has been attributed to exchange interaction.
Another signature related to electron-electron interactions in TMDCs is the presence of odd-to-even filling factor transitions in the magnetoresistance \cite{movva_density-dependent_2017, pisoni_interactions_2018, pisoni_absence_2019}, which has been attributed to a density-dependent Landé g factor.
The filling factor transition can be observed also in our experiments, as shown by the filling factors reported in Fig.~\ref{Fig:Fig2}(a). 
In light of these findings, we attribute our observation of a decreasing density in the top layer upon passing $\vbg^{(\alpha)}$ to electron-electron interactions.

We calculate the residual density (that is, the density in the lower two layers) according to $n_\mathrm{diff} = n_\mathrm{s}-n_\mathrm{TL}$ to obtain the black solid line in Fig.~\ref{Fig:Fig4}(b).
We find that the residual density associated with the two lower layers first increases linearly with $\vbg$, starting from $\vbg^{(\alpha)}$ up to $\vbg^{(\beta)}$ (black dashed line). 
The rate of this increase is larger than the rate of increase of the total density $n_\mathrm{s}$ due to the redistribution of carriers from the top layer. 
Then the slope of the black line decreases abruptly at $\vbg^{(\beta)}$, converging toward the slope of the total density $n_\mathrm{s}$ (red dashed line), while $n_\mathrm{TL}$ barely changes any more. This indicates that the charge redistribution from the top to the bottom layers has come to an end for $\vbg>\vbg^{(\beta)}$.

\begin{figure}[tb]
\begin{center}
\includegraphics[scale=1]{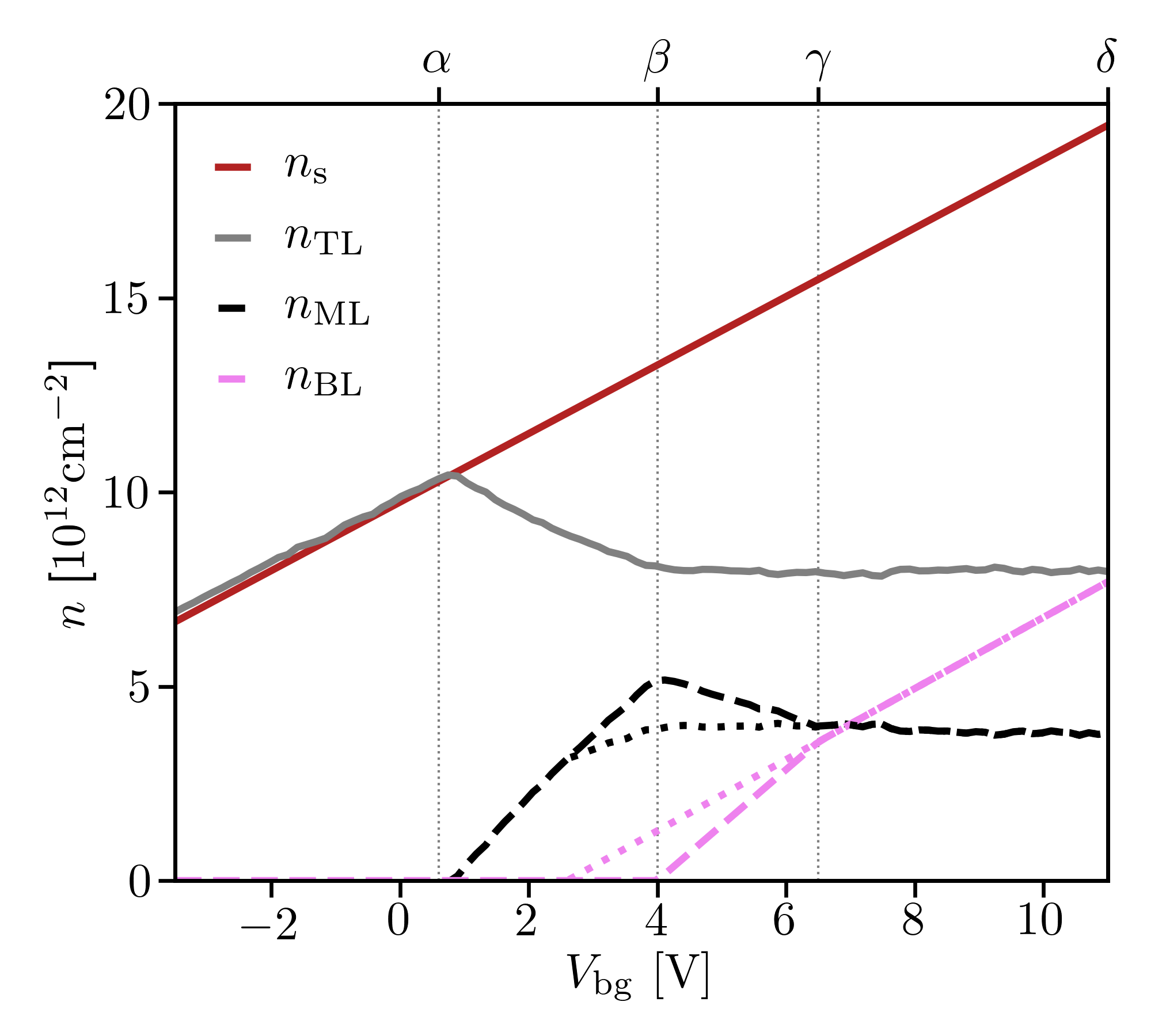}
\caption{Proposed evolution of the densities in the bottom and middle layers as a function of $\vbg$ at $\vtg=\SI{12}{V}$. 
The electron density in the outermost layers are given by $n_\mathrm{TL}=n_1 + n_2$ and $n_\mathrm{BL}=n_3 + n_4$, and the density in the middle layer is obtained by $n_\mathrm{s}-n_\mathrm{TL}-n_\mathrm{BL}$. 
The dotted lines are obtained by linearly extrapolating the density $n_\mathrm{BL}$, while the dashed line presents a second scenario, based on the hypothetical onset of the band at $V_\mathrm{bg}^{(\beta)}$. The letters in the top axis refer to the features observed in the resistance [Fig.~\ref{Fig:Fig1}(b)].}
\label{Fig:Fig7}
\end{center}
\end{figure}

We now focus on the second Landau fan observed in Fig.~\ref{Fig:Fig2} (violet lines).
In Fig.~\ref{Fig:Fig5}(a) we plot $dR_{xx}/d\vbg$ as a function of $B$ and $\vbg$ at $\vtg=\SI{10}{V}$. 
This plot is like Fig.~\ref{Fig:Fig2}(b) for $\vbg>\SI{6}{V}$, but taken at a slightly smaller $\vtg$.
Reducing the voltage $\vtg$ allows us to resolve additional frequencies that were not clearly visible at $\vtg=\SI{12}{V}$.
The FFT is shown in Fig.~\ref{Fig:Fig5}(b), where we highlight three frequencies (arrows).
Analyzing the Fourier spectra for $\vtg$ between $\SI{8}{V}$ and $\SI{12}{V}$, we determine the $\vbg$ dependence of frequencies $f_3$ and $f_4$, which is shown in Fig.~\ref{Fig:Fig5}(c).
These two frequencies are seen to be independent of $\vtg$.
This behavior is evident also from the gate-gate map at $B=\SI{9}{T}$ in Fig.~\ref{Fig:Fig2}(c), where the SdH oscillations with frequency $f_3$ appear as vertical lines, and consistent with our notion of a strong top-layer population.

We determine the electron densities $n_3$ and $n_4$ associated with the frequencies $f_3$ and $f_4$ as we did for the top layer.
At this point we have to assume a degeneracy factor for the densities in these two bands.
Among the two possibilities (sixfold and twofold degeneracy for $Q$ and $K$ valleys, respectively), only a twofold degeneracy leads to electron densities comparable with the residual density $n_\mathrm{diff}$ (shown in Fig.~\ref{Fig:Fig4}).
If we assumed a six fold degeneracy, the density $n_3+n_4$ would be much larger than $n_\mathrm{diff}$.
From the odd-to-even filling factor transition observed in Fig.~\ref{Fig:Fig2}(a), we conclude that $g_i \geq 2$.
Assuming $g_{i}=2$ ($i=3,4$), the $\vbg$ dependence of $n_3+n_4$ yields a gate capacitance of $C_\mathrm{bg}\approx \SI{147}{nF/cm^2}$, which is in agreement with the expected capacitance $\SI{145(5)}{nF/cm^2}$ between the bottom gate and the bottom layer. 
This back-gate capacitance in the positive $\vbg$ range is slightly larger than in the single layer regime at negative $\vbg$, indicating that the conducting plane is closer to the bottom gate, as expected. 

The resulting densities are shown in Fig.~\ref{Fig:Fig6} together with the densities $n_1$ and $n_2$ of the top layer, the Hall density $n_\mathrm{H}$ and the total sheet density $n_\mathrm{s}$ determined from the capacitance model.
Remarkably, there is a symmetry between the densities $n_1$ and $n_2$ of the top layer and the densities $n_3$ and $n_4$ at the highest bottom-gate voltage [$V^{(\delta)}$], such that $n_3\approx n_1$ and $n_4\approx n_2$. 
However, summing up all four densities does not yield the total sheet electron density $n_\mathrm{s}$, but there seems to be a latent density invisible in the SdH analysis (see black solid line in Fig.~\ref{Fig:Fig6}). 
However, the total increase of $n_3+n_4$ accounts for the total increase of $n_\mathrm{s}$ for $\vbg>\SI{9}{V}$. 
This indicates that the latent density is constant in this gate voltage range, i.e., it is well screened from the influence of the bottom gate. 
In addition, changing the top-gate voltage from $\SI{9}{V}$ to $\SI{12}{V}$ does not significantly influence the latent density, suggesting that this density is also well screened from the influence of the top gate (at least in this limited voltage range).
Based on these observations, we draw the conclusion that the latent density must reside in the middle layer (hence, the label $n_\mathrm{ML}$ in Fig.~\ref{Fig:Fig6}).
The densities $n_3$ and $n_4$ have a similar gate dependence, and no saturation is observed even at large $\vbg$.
Therefore, the densities $n_3$ and $n_4$ correspond to the two spin-valley locked bands of the bottom layer.

This conclusion is also justified by the symmetry of our layered device in the out-of-plane direction.
The top and bottom layers experience the same dielectric environment, i.e., both of them have a hBN layer on one side and an \mos layer on the other side. 
Thus, we would expect that they have similar electronic properties. 
In addition, the voltage range considered here is very close to the condition of zero displacement field, where we would expect a symmetric charge distribution along the out-of-plane direction. 
This is indicated in the top inset of Fig.~\ref{Fig:Fig6} by the configuration at the very right. 
The schematic on the very left of this inset shows the situation at negative $\vbg$, where only the top layer is populated.

The discussion of the density evolution of $n_3$, $n_4$, and $n_\mathrm{ML}$ in the range of $\vbg$, where no SdH data is available for these densities remains speculative.
In the top inset of Fig.~\ref{Fig:Fig6} we propose a possible scenario in which the layers are sequentially populated with electrons as we tune the bottom gate from negative to positive. 
Figure~\ref{Fig:Fig7} presents the density in the three layers according to this interpretation.
The deviation of $n_\mathrm{TL}$ from $n_\mathrm{s}$ at $\vbg^{(\alpha)}$ coincides with the increase of the zero-field resistance $R_{xx}$ observed in Fig.~\ref{Fig:Fig1}(b).
These two signatures are the experimental evidence of a second layer being populated.
However, the linear interpolation of $n_\mathrm{BL}$ does not coincide with $\vbg^{(\alpha)}$ (see pink dotted line in Fig.~\ref{Fig:Fig7}). 
Whether the middle or the bottom layer is populated depends on the penetration field.
Owing to the finite density of states of the top layer, some field penetration occurs \cite{luryi_quantum_1988}, resulting in different onsite potential for the middle and bottom layers at $\vtg>0$ and $\vbg=0$. 
In this case, upon increasing $\vbg$,  the middle layer might be occupied at lower $\vbg$ than the bottom layer; thus, $\vbg^{(\alpha)}$ corresponds to the occupation of a band in the middle layer.

We hypothesize that the peak in the resistance at $\vbg^{(\beta)}$ in Fig.~\ref{Fig:Fig1}(b) corresponds to the onset of an additional band.
This voltage is close although not identical to the linearly extrapolated onset of $n_\mathrm{BL}$.
The reason could be related to interaction effects, which lead to a larger rate $dn_\mathrm{BL}/d\vbg$ in the low-density range, as observed in Fig.~\ref{Fig:Fig4}(b).
Therefore, we assume that the onset of the density $n_\mathrm{BL}$ occurs at $\vbg^{(\beta)}$.
In this case, the densities in the middle and bottom layers are described by the dashed lines in Fig.~\ref{Fig:Fig7}.
At the onset of the band in the bottom layer ($\vbg^{(\beta)}$), the density in the middle layer decreases, as observed for $n_\mathrm{TL}$ at $\vbg^{(\alpha)}$.
Concomitantly, the enhanced density of states due to the additional band provides additional screening, and the density in the top layer ($n_\mathrm{TL}$) remains constant for $\vbg>\vbg^{(\beta)}$.

Following this hypothesis, the peak in the resistance at $\vbg^{(\gamma)}$ corresponds to the onset of a second band (the upper SO split band) of the bottom layer. 
In fact, at this voltage, the density in the bottom layer is $\SI{3.6E12}{cm^{-2}}$, which is the same density reported for the population of the upper SO split band in bilayer \mos \cite{pisoni_absence_2019}.
Again, the additional density of states provides additional screening, and the density in the middle layer remains roughly constant.
For $\vbg>\vbg^{(\gamma)}$, the rate of change $dn_\mathrm{BL}/d\vbg$ accounts for the change in the total sheet density, while the densities in the top and middle layers are almost completely screened by the electrons in the bottom layer.

\begin{figure*}[tb]
\begin{center}
\includegraphics[scale=1]{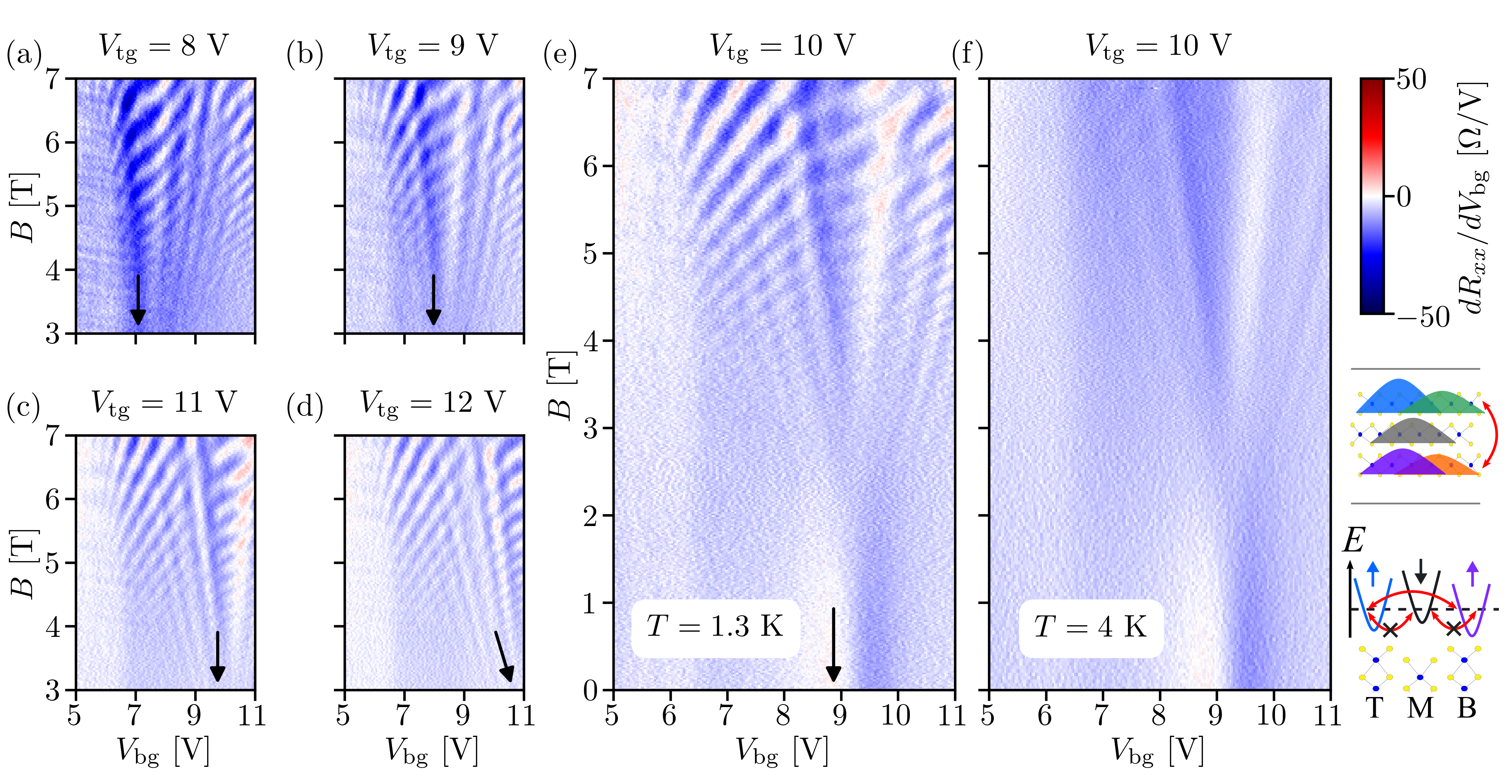}
\caption{(a)-(e) $dR_{xx}/d\vbg$ as a function of $B$ and $\vbg$ for $\vtg$ between $\SI{8}{V}$ and $\SI{12}{V}$ at $T=\SI{1.3}{K}$. The black arrows point toward $D=0$.
(f) $dR_{xx}/d\vbg$ as a function of $B$ and $\vbg$ at $\vtg=\SI{10}{V}$ at $T=\SI{4}{K}$.
On the right panel, there are two schematics showing scattering between the outermost layers, while the middle layer is completely decoupled due to a different spin polarization. The letters T, M, and B stand for top, middle, and bottom layers.}
\label{Fig:Fig8}
\end{center}
\end{figure*}

In our experiments, we do not observe any evidence for $Q$ valley states in 3L \mos (as in bilayer \mos \cite{pisoni_absence_2019}).
The lack of $Q$-valley states might be related to the encapsulating hBN used for fabricating our devices \cite{rickhaus_gap_2019}.
DFT calculations usually assume vacuum surrounding the \mos layers.
According to these calculations, the $Q$-valley states inherit the $p_z$ orbital-character of the sulfur atoms, and these orbitals are directed out of plane. 
The hybridization of these orbitals is the prime reason for the shift in energy of the $Q$ valleys as we add single \mos layers one-by-one starting from  a monolayer.
It is therefore likely that a large band gap insulator like hBN might affect the band structure of \mos close to the $Q$ points.
The orbital hybridization of the outer sulfur atoms with the orbitals of the atoms in the hBN layer might change the band edge energy of the $Q$ valley of the outer \mos layers \cite{falko_privcom_2020}.
$Q$ valley states of the outermost layers do not hybridize with those of the middle layers, such that no downward shift of the $Q$-valleys is observed in encapsulated 3L \mos.
On the other hand, $K$-valley states (mainly with in-plane $d$ orbital character of the transition metal atom) are not strongly affected by the dielectric environment, as they are almost not affected by the number of \mos layers. 
As a result $K$-valley states lie lower in energy than the $Q$ valleys, as it is the case for monolayer \cite{pisoni_interactions_2018} and bilayer \mos \cite{pisoni_absence_2019}. 

\subsection{Interlayer scattering}

Finally, we would like to discuss feature $\delta$ in Fig.~\ref{Fig:Fig1}(b).
To overcome the Schottky barrier at the interface between the metallic contacts and the \mos, we must apply a large $\vtg$. 
Therefore, most of our measurements are performed at finite displacement field ($D$).
However, in the data presented in Fig.~\ref{Fig:Fig1}(c), we explore a large $D$ range, including $D=0$.
Along the zero-displacement field line, where the electron densities in the top and bottom layers are the same (in agreement with the previous SdH analysis), we observe a peak in $R_{xx}(B=0)$.
In perpendicular magnetic field we see that Landau fanlike features emerge from the $D=0$ line, shown in Figs.~\ref{Fig:Fig8}(a)-\ref{Fig:Fig8}(e), where $dR_{xx}/dV_\mathrm{bg}$ is plotted as a function of $(B, \vbg)$ for $\vtg$ between $\SI{8}{V}$ and $\SI{12}{V}$.

Similar features have been observed in twisted bilayer $\mathrm{WSe_2}$ \cite{wang_magic_2019}, where for a specific range of twist angles, electronic interactions lead to the opening of a small band gap in the spectrum (Mott-like insulator state). 
Here, we exclude the presence of a gap for two reasons. 
First, the Hall resistance does not show any change of sign. 
Second, the magnetoresistance oscillates with a very low frequency [see black arrow in Fig.~\ref{Fig:Fig5}(b)].
The corresponding electron density would be $\leq\SI{5E11}{cm^{-2}}$, which is suspicious, given the fact that we never observed SdH oscillations for densities lower than $\leq\SI{1.2E12}{cm^{-2}}$ in \mos devices.
We note that the frequency of these oscillations is proportional to the difference $|n_\mathrm{TL} - n_\mathrm{BL}|$, suggesting magneto-interband scattering as a possible explanation \cite{raikh_magnetointersubband_1994}.
At finite magnetic field, the resistance is expected to increase whenever the LLs of two bands are in resonance, leading to magneto-oscillations periodic in $B^{-1}$ and with a frequency proportional to the energy offset of the two bands (i.e., proportional to the difference of the densities in the two bands), as in our data.

It is noteworthy that interlayer tunnel coupling has not been observed in bilayer \mos \cite{pisoni_absence_2019} where the two layers are directly in contact with each other.
Therefore, it seems unlikely at first glance that tunnel coupling increases in 3L \mos, where the middle layer has to mediate the tunneling.
Instead, it is more likely that electron scattering mediated by impurities is the underlying mechanism.
When the bands are energetically in resonance, a scattering channel opens, and the resistance increases.
This effect has not been observed in bilayer \mos, where the adjacent layers of the 2H polytype are rotated by \SI{180}{\degree} with respect to each other.
As a consequence, the spin polarization within the same $K$ valley in two adjacent layers is opposite.
Therefore, interlayer scattering in bilayer \mos would require either a spin flip or a large momentum transfer (on the scale of the first Brillouin zone), thus suppressing such a scattering mechanism.
On the other hand, the outermost layers in 3L \mos are equally oriented. As a result, the respective $K$ valleys have the same spin polarization, as shown by the schematics in the lower right panel of Fig.~\ref{Fig:Fig7}.
In this case, the scattering between the layers is not suppressed, leading to the observed resistance peak at zero displacement field and magneto-interlayer oscillations \cite{falko_privcom_2020}. 

To verify our hypothesis, we performed measurements as a function of temperature.
Magneto-interlayer oscillations have a weak temperature dependence and should persist at higher temperatures than SdH oscillations \cite{raikh_magnetointersubband_1994}.
Figure~\ref{Fig:Fig8}(f) shows $dR_{xx}/d\vbg$ as a function of $(B,\vbg)$ at $\vtg=\SI{10}{V}$ and $T=\SI{4}{K}$.
At this temperature, SdH oscillations are completely suppressed due to the large electron effective mass ($m^{*}\approx 0.6-0.8 m_0$ \cite{pisoni_interactions_2018, pisoni_absence_2019}, $m_0$ being the bare electron mass), whereas the low frequency magneto-oscillations persist.
The insensitivity to temperature confirms that magneto-interlayer scattering is the underlying mechanism of these magneto-oscillations, whereas their frequency demonstrates that electron scattering takes place between the outermost layers.

Despite the inverted spin-valley symmetry of adjacent \mos layers, we observe signatures of interlayer scattering close to $V_\mathrm{bg}^{(\alpha)}$, when the top layer is heavily populated by electrons and the middle layer starts to be occupied. 
At the onset of the population of the middle layer, we observe two effects: charge redistribution among the two layers and an increase in the resistance [see Fig.~\ref{Fig:Fig1}(b)].
The reduction of the charge density in the top layer is not enough to quantitatively describe the resistance increase, even if we assume that the middle layer is in the mobility gap.
Therefore, interlayer scattering must be considere to describe the increasing resistance at $V_\mathrm{bg}^{(\alpha)}$.
The underlying mechanism is not clear yet.
A possible candidate is the presence of defect states at the bottom of the conduction band in the middle layer, which could provide enough momentum to induce interlayer scattering between adjacent layers. 
This effect is only relevant close to the band edges, where the electron density and the mobility are too low for observing magneto-interband oscillations.

\section{Conclusions}

In this paper we investigated electron transport through 3L \mos to understand the alignment, the population, and the coupling of conduction bands in the different layers. 
For this purpose, we employed a dual-gated 3L \mos device with bottom metallic contacts and performed four-terminal magnetotransport measurements. 
By analyzing the Fourier spectra of the SdH oscillations, we were able to determine the densities in the individual layers.
We experimentally demonstrated that the conduction band minima of 3L \mos lie at the corners of the hexagonal Brillouin zone ($\pm K$ points), as testified by the observed twofold LL degeneracy.
Filling the layers with electrons, we observed interaction effects, such as negative compressibility, and nonmonotonic changes in the zero-field resistance with total carrier density.
In contrast to previous reports, we did not observe any population of the $Q$ valleys.
In addition, we measured a peak in the resistance at zero displacement field, when the densities in the two outermost layers are the same. We interpret this observation as a consequence of interlayer scattering.
This interpretation is confirmed by the observed magneto-interband oscillations with a frequency proportional to the difference of the densities in the outermost layers.

\section*{Acknowledgments}
We thank Rebekka Garreis, Annika Kurzmann, and Chuyao Tong for fruitful discussions.
We thank Peter Märki, Thomas Bähler, as well as the FIRST staff for their technical support.
We acknowledge financial support from the European Graphene Flagship and the Swiss National Science Foundation via NCCR Quantum Science and Technology.
K. W. and T. T. acknowledge support from the Elemental Strategy Initiative conducted by the MEXT, Japan, Grant No. JPMXP0112101001,  JSPS KAKENHI Grant No. JP20H00354 and the CREST(JPMJCR15F3), JST.

%

\end{document}